\documentclass{aa}
\usepackage{psfig}
\begin{document}
\thesaurus{12.03.4; 12.12.1; 11.06.1}
\title{Non radial motions and the shapes and the abundance of clusters of
galaxies}
\author{A. Del Popolo\inst{1,2}, M. Gambera\inst{1,3}}
\offprints{M. Gambera (mga@sunct.ct.astro.it)}
\institute{$^1$ Istituto di Astronomia dell'Universit\`a di Catania, 
Viale A.Doria, 6 - I 95125 Catania, Italy \\
$^2$ Dipartimento di Matematica, Universit\`{a} Statale di Bergamo, Piazza Rosate, 2 - I 24129 Bergamo, Italy \\
$^3$ Osservatorio Astrofisico di Catania and CNR-GNA, 
Viale A.Doria, 6 - I 95125 Catania, Italy 
}
\date{}
\maketitle

\begin{abstract}
We study the effect of non-radial motions on the mass function, the {\it velocity
dispersion function} (hereafter VDF) and on the shape of clusters of galaxies using
the model introduced in Del Popolo \& Gambera (1998a,b; 1999). The mass
function of clusters, obtained using the quoted model,
is compared with the statistical
data by Bahcall \& Cen (1992a,b) and Girardi et al. (1998),
while the VDF is compared with the Center
for Astrophysics (hereafter CfA) data by 
Zabludoff et al. (1993) for local clusters and those of Mazure et al. (1996) and
Fadda et al. (1996). In both cases the model
predictions are in good agreement with the observational data showing once more
how non-radial motions can reduce many of the discrepancies between 
Cold Dark Matter (hereafter CDM) 
model predictions and observational data. Finally we study the effect of non-radial motions
on the intrinsic shape of clusters of galaxies showing that non-radial
motions produce clusters less elongated with respect to CDM model in agreement
with de Theije et al. (1995, 1997) results.
\keywords{cosmology: theory-large scale structure of Universe - galaxies: formation}
\end{abstract}

\section{Introduction }

At its appearance the CDM model
contributed to obtain a better
understanding of the origin and evolution of the large scale structure in the
Universe (White et al. 1987; Frenk et al. 1988; Efstathiou 1990; Ostriker
1993).
The principal assumptions of the {\it standard} CDM (SCDM) model (see also
Liddle \& Lyth 1993) are: 
\begin{itemize}
\item a flat Universe dominated by weakly
interacting elementary particles having low velocity dispersion at early
times. The barionic content is determined by the standard big bang
nucleosynthesis model (Kernan \& Sarkar 1996; Steigman 1996; Olive 1997; 
Dolgov 1997); 
\item critical matter density; 
\item expansion rate given by $ h = 0.5$; 
\item a scale invariant and adiabatic spectrum with a spectral index, 
$ n \equiv 1$; 
\item the condition required by observations, that the fluctuations
in galaxy distribution, $(\delta \rho / \rho)_{\rm g}$ , are larger than the
fluctuations in the mass distribution, $(\delta \rho / \rho)_{\rho} $ by 
a factor $ b > 1$. 
\end{itemize}
If this last assumption is not introduced, the pairwise 
velocity dispersion is larger then that deduced
from observations and the galaxy correlation function is steeper than that
observed (Davis et al. 1985). After the great success of the model in the 80's, a closer inspection of the model has shown a series of deficiencies, namely:
\begin{itemize}
\item the strong
clustering of rich clusters of galaxies, 
$ \xi_{\rm cc}(r) \simeq (r / 25h^{-1}{\rm Mpc})^{-2}$, 
far in excess of CDM predictions (Bahcall \&
Soneira 1983);
\item the overproduction of clusters abundance. 
Clusters abundance is a useful test for models of galaxy formation. This
is connected to three relevant parameters: the mass function,
the VDF and the temperature
function.
Using N-body simulations, Jing et al. (1994) studied the mass function
of rich clusters at $z=0$ for the CDM model concluding, if the density spectrum is
normalized to the Cosmic Background Explorer (hereafter COBE) (Smoot et al. 1992) quadrupole $Q_{COBE}=6.0 \times 10^{-6}$,
that the mass function is higher than the observed one by Bahcall \& Cen (1992a,b).
Bartlett \& Silk (1993) come to a similar conclusion using
the Press-Schechter (1974) formula. They 
found that the CDM model with the COBE normalization
produces a temperature function of clusters higher than that given by
the observations by
Edge et al. (1990) and by Henry \& Arnaud (1991);
\item the
conflict between the normalization of the spectrum of the perturbation which
is required by different types of observations; in fact, 
the normalization obtained from
COBE data (Smoot et al. 1992) on scales of the order of $10^3$ Mpc 
requires $\sigma_8=0.95\pm 0.2$, where
$\sigma_8$ is the
rms value of $\frac{\delta M}{M}$ in a sphere of $8 h^{-1}$Mpc.
Normalization on scales $10 \div 50$Mpc obtained from QDOT
(Kaiser et al. 1991) and POTENT (Dekel et al. 1992) requires that $\sigma _8$
is in the $0.7 \div 1.1$, range which is compatible with COBE normalization
while the observations of the pairwise velocity dispersion of galaxies on
scales $r\leq 3$ Mpc seem to require $\sigma _8 \leq 0.5$. 
\item 
the incorrect scale dependence of the galaxy correlation
function, $\xi (r)$, on scales
$10 \div 100$ $h^{-1} {\rm Mpc}$, having $\xi (r)$ too
little power on the large scales compared to the power on smaller scales
(Maddox et al. 1990; Saunders et al. 1991; Lahav et al. 1989;
Peacock 1991; Peacock \& Nicholson 1991);  
the APM survey (Maddox et al. 1990), giving the galaxy angular
correlation function, the 1.2 Jy IRAS power spectrum, the QDOT survey
(Saunders et al. 1991), X-ray observations (Lahav et al. 1989) and radio
observations (Peacock 1991; Peacock \& Nicholson 1991) agree with the quoted
conclusion. As shown in studies of galaxy clustering on large scales
(Maddox et al. 1990; Efstathiou et al. 1990b; Saunders et al. 1991) the
measured rms fluctuations within spheres of radius $20h^{-1}$Mpc have
value $ 2 \div 3$ times larger than that predicted by the CDM model.
\end{itemize}
Several alternative models have been proposed in order to solve
the quoted problems (Peebles 1984; Shafi \& Stecker 1984; 
Valdarnini \& Bonometto 1985; Bond et al. 1988; Schaefer et al. 1989; 
Holtzman 1989; Efstathiou et al. 1990a; Turner 1991; 
Schaefer 1991; White et al. 1993a; Shaefer \& Shafi 1993; Holtzman \& Primack 1993; Bower et al. 1993). Most of them propose 
in some way a modification of the primeval spectrum of perturbations.  
In two previous papers (Del Popolo \& Gambera 1998a; 1999) we showed
how, starting from a CDM spectrum and taking into account
non-radial motions, at least the problem of the clustering of clusters of
galaxies (first point above) and 
the problem of the X-ray temperature 
(second problem quoted above) can be considerably reduced. \\
In this paper
we extend the model to two other tests of the abundance of clusters:
the mass function and the VDF, permitting to estimate
the expected number density of
clusters within a given range in mass and velocity, respectively.
We also study the effect of non-radial motions on the shapes of galaxy
clusters.
In recent papers
(de Theije et al. 1995; de Theije et al. 1997) it has been shown that
clusters of galaxies are more nearly spherical and more centrally
condensed than the predictions of CDM models with $\Omega=1$. \\
As we shall see, non-radial
motions have the effect to produce more spherical clusters and are
able to reconcile the CDM with $\Omega=1$ predictions on clusters elongations
with observations. \\
In Sect. ~2 we shall use the same model introduced by
Del Popolo \& Gambera (1998a,b; 1999)
to take into account non-radial motions, arising from the tidal
interaction of the protoclusters with the neighbouring protostructures, and
we shall compare the mass function calculated using the CDM model,
taking into account non-radial motions, with the observed mass function
obtained by Bahcall \& Cen (1992)
and Girardi et al. (1998).
In Sect. ~3 we repeat the calculation for
the VDF and compare the theoretical VDF with the CfA data by 
Zabludoff et al. (1993)
and with the data by Mazure et al. (1996) and
Fadda et al. (1996). 
In Sect. ~4 we study the effect of non-radial motions
on the ellipticity of clusters and finally in Sect. ~5 we give our conclusions.
 




\section{Non-radial motions and the mass function}

One of the most important constraints that a model for large-scale
structure must overcome is that of predicting the correct number density of
clusters of galaxies. This constraint is crucial for several reasons.
According to the gravitational instability scenario, galaxies and clusters
form where the density contrast, $\delta$, is large enough so that the surrounding
matter can separate from the general expansion and collapse. Consequently
the abundance of collapsed objects depends on the amplitude of the
density perturbations. In the CDM model these latter follow a Gaussian
probability distribution and their amplitude on a scale $R$
is defined by $\sigma(R)$, the r.m.s. value of $\delta$,
which is related to the power
spectrum, $P(k)$. In hierarchical models of structure formation, like
CDM, $\sigma(R)$ decreases with increasing scale, $R$, and consequently
the density contrast required to form large objects, like clusters of galaxies,
rarely occurs. The present abundance of clusters is then extremely
sensitive to a small change in the spectrum, $P(k)$. Moreover, the rate
of clusters evolution is strictly connected to the density parameter, $\Omega_0$.
Then, clusters abundance and its evolution are a probe of
$\Omega_0$ and $P(k)$
and can be used to put some constraints on them.   \\
The abundance of clusters of galaxies, 
together with the mass distributions in galaxy halos and in rich clusters
of galaxies, the peculiar motions of galaxies, the spatial
structure of the microwave background radiation is one of the most
readily accessible observables which probes the mass distribution
directly.

The most accurate way of assessing the cluster abundance is via numerical
simulations. However, there is an excellent analytic alternative, Press \&
Schechter's theory (Press \& Schechter 1974; Bond et al. 1991).
Brainerd \& Villumsen (1992) studied the
CDM halo mass function using a hierarchical particle mesh code. From this
last work  it results that the Press \& Schechter formula fits the results 
of the simulation up
to a mass of ~10 times the characteristic 1$\sigma$ fluctuation mass,
$M_{\ast}$, being $M_{\ast} \simeq 10^{15} b^{-6/(n_{l}+3)} M_{\odot}$,
where $b$ is the
bias parameter 
and $n_{l}$ is the local slope of the power spectrum.
For the case of critical-density universes, N-body simulations (Lacey
\& Cole 1994) have been shown to be in extremely well agreement with
Press \& Schecter's theory. \\
Press-Schechter's theory states that the
fraction of mass in gravitationally bound systems larger than a mass, $M$,
is given by the fraction of space in which the linearly evolved
density field, smoothed on the mass scale $M$, exceeds a threshold
$\delta_{\rm c}$:
\begin{equation}
F(>M)=\frac{1}{2} erfc \left(\frac{\delta_{\rm c}}{\sqrt{2}
\sigma(R_f,z)}   \right)
\label{eq:press}
\end{equation}
The fraction of the mass density in non-linear objects of mass M
to M+d M is given by differentiating Eq. (\ref{eq:press}) with respect
to mass:
\begin{equation}
n(M,z) dM= -\left(\frac{2}{\pi}\right)^{1/2}
\frac{\rho_{\rm b}}{M} \cdot \frac{\delta_{\rm c}}{\sigma^2}
\exp(-\frac{\delta_{c^2}}{2 \sigma^2}) \frac{{\rm d} \sigma}{{\rm d} M}
{\rm d} M
\label{eq:press1}
\end{equation}
where $\rho_{\rm b}$ is the comoving background density, $R_{\rm f}$ is the
comoving linear scale associated with $M$,
$R_f=\left(\frac{3 M}{4 \pi \rho_{\rm b}}\right)^{1/3}$.
Press-Schechter's result predicts that only half of the mass of the Universe
ends up in virialized objects but in particular cases this problem can
be solved (Peacock \& Heavens 1990; Cole 1991; Blanchard et al. 1992). \\
The mass variance present in Eq.~(\ref{eq:press})
can be obtained once a spectrum, $P(k)$, is fixed:
\begin{equation}
\sigma ^2(M)=\frac 1{2\pi ^2}\int_0^\infty {\rm d}kk^2P(k)W^2(kR) 
\label{eq:ma3}
\end{equation}
where $W(kR)$ is a top-hat smoothing function:
\begin{equation}
W(kR)=\frac 3{\left( kR\right) ^3}\left( \sin kR-kR\cos kR\right) 
\label{eq:ma4}
\end{equation}
and the power spectrum $P(k)=Ak^nT^2(k)$ is fixed giving the transfer
function $T(k):$
\begin{eqnarray}
T(k) &=& \frac{[\ln \left(1+18k \sqrt{a}\right)]}{18 \sqrt{a}}
\cdot [1+1.2 k^{1/2}-27k+ \nonumber \\
& + &  347 (1-\sqrt{a}/5) k^{3/2}-18 (1-0.32 a^2) k^2]^{-2}
\label{eq:ma5}
\end{eqnarray}
(Klypin et al. 1993), where $A$ is the normalization constant,
$a=(1+z)^{-1}$ is the expansion parameter and $k$ is
the wave-number measured in units of ${\rm Mpc}^{-1}$. This spectrum is valid for
$k< 30 {\rm Mpc}^{-1}$ and $z<25$. The accuracy (maximum deviation) of
the spectrum is $5\%$. It is more accurate than Holtzman's (1989) spectrum,
used by Jing et al. (1994) and Bartlett \& Silk (1993) to calculate
the mass function and the X-ray temperature function of clusters, respectively.
The spectrum is lower by $20\%$ on cluster mass scales than 
Holtzman's (1989). The spectrum was normalized to the COBE quadrupole
$Q_2=17 \mu$ K, corresponding to
$\sigma_8=0.66$.
As shown by Bartlett \& Silk (1993) the X-ray distribution
function, obtained using a standard CDM spectrum, over-produces the clusters
abundances data obtained  by Henry \& Arnaud (1991) and by Edge et al.
(1990).
This has lead some authors (White et al. 1993b)
to cite the cluster abundance
as one of the strongest pieces of evidence against the standard
CDM model when the model is normalized so as to reproduce the
microwave background anisotropies as seen by the COBE satellite
(Bennett et al. 1996; Banday et al. 1996; G\'orsky et al. 1996;
Hinshaw et al. 1996). \\
The
discrepancy can be reduced taking into account the non-radial motions that
originate when a cluster reaches the non-linear regime as follows.
A fundamental role in Press-Schechter's theory 
is played by the value of $\delta_{\rm c}$. This value
is quite dependent on the choice of smoothing window used to obtain the
dispersion (Lacey \& Cole 1994). Using a top-hat window function
$\delta_{\rm c}=1.7 \pm 0.1$ while for a Gaussian window the threshold is
significantly lower. In a non-spherical context the situation is more
complicated. Considering the collapse along all the three axes the threshold
is higher, whereas the collapse along the first axis (pancake
formation) or the first two axes (filament formation) corresponds
to a lower threshold (Monaco 1995). The threshold, $\delta_{\rm c}$,
does not depend on the background cosmology. \\
As shown by Del Popolo \& Gambera (1998a; 1999), if non-radial
motions are taken into account, the threshold $\delta_{\rm c}$
is not constant but is function of mass, $M$ (Del Popolo 
\& Gambera 1998a; 1999):
\begin{equation}
\delta _{\rm c}(\nu )=\delta _{\rm co}\left[ 1+
\int_{r_{\rm i}}^{r_{\rm ta}}  \frac{r_{\rm ta} L^2 \cdot {\rm d}r}{G M^3 r^3}%
\right]
\label{eq:ma7} 
\end{equation}
where $\delta _{\rm co}=1.68$ is the critical threshold for a spherical model,
$r_{\rm i}$ is the initial radius, $r_{\rm ta}$ is the turn-around radius and
$L$ the angular momentum.
In terms of the Hubble constant, $H_0$, the density parameter at current
epoch, $\Omega_0$, the expansion parameter $a$ and
the mean fractional density excess inside a shell of a given radius,
$\overline{\delta}$ Eq. (\ref{eq:ma7}) can be written as (Del Popolo 
\& Gambera 1998a; 1999):
\begin{equation}
\delta _{\rm c}(\nu )=\delta _{\rm co}\left[ 1+\frac{8G^2}{\Omega
_o^3H_0^6r_{\rm i}^{10}\overline{\delta} (1+\overline{\delta} )^3}
\int_{a_{\rm min}}^{a_{\rm max }}\frac{L^2 \cdot {\rm d}a}{a^3}%
\right]
\label{eq:ma8} 
\end{equation}
where $a$ is the expansion parameter,
and $a_{\rm min}$ its value corresponding to 
$r_{\rm i}$.
The mass dependence of the threshold parameter,
$ \delta_{\rm c}(\nu)$, was obtained
in the same way sketched in Del Popolo \& Gambera (1999): 
we calculated the binding radius, $r_{\rm b}$, of the shell 
using Hoffmann \& Shaham's criterion (1985): 
\begin{equation}
T_{\rm c}(r, \nu) \leq t_{0}
\label{eq:ma9}
\end{equation}
where $T_{\rm c}(r,\nu)$ is the calculated time of collapse of a shell and $
t_{\rm o}$ is the Hubble time. We found a relation between $ \nu$ and $M$ through
the equation $ M=4 \pi\rho_{\rm b} r^{3}_{\rm b}/3$. 
\begin{figure}[ht]
\psfig{file=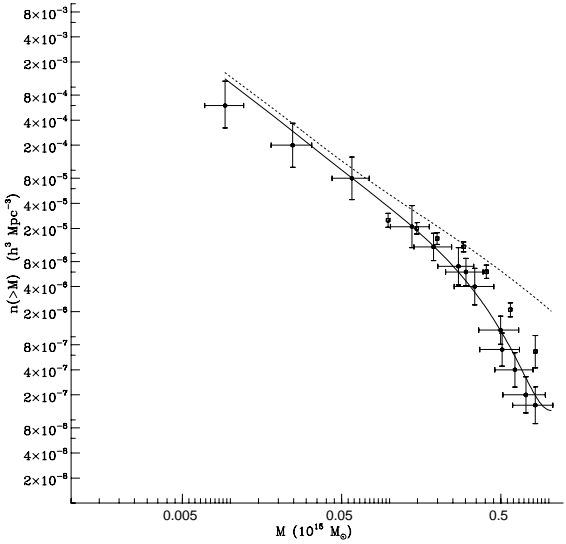,width=8.cm}
\caption{Cumulative mass function calculated using a CDM model without
taking into account non-radial motions (dashed line) and taking account
non-radial motions (solid line) compared with Bahcall \& Cen (1992) data (full dots)
and
with that of Girardi et al. (1998) (open squares).}
\end{figure}
We so obtained $ \delta_{\rm c}[\nu(M)]$. We obtained the
total specific angular momentum, $h(r,\nu )=L(r,\nu)/M_{\rm sh}$, 
acquired during expansion,
in the same way sketched in Del Popolo \& Gambera (1998a; 1999). 
Taking account that $\delta_{\rm c}$ depends on $M$ Eq. (\ref{eq:press1}) 
becomes:
\begin{equation}
n(M,z) {\rm d}M=\left(\frac{2}{\pi}\right)^{1/2} \frac{\rho}{M} 
\frac{\exp(-\frac{\delta_{\rm c}^2)}{2 \sigma^2}}{\sigma^2}
\left(\sigma \frac{{\rm d} \delta}{{\rm d} M} -\delta_{\rm c} \frac{{\rm d} \sigma}{{\rm d} M} \right) {\rm d}M
\label{eq:press2}
\end{equation}
The result of the calculation is shown in
Fig. 1.
Here, the mass function of clusters, derived using
a CDM model with $\Omega_0=1$, $h=1/2$ normalized to
$Q_{\rm COBE} = 17 \mu$ K
and taking into account
non-radial motions (solid line), is compared with 
the statistical data by Bahcall \& Cen (1992) (full dots)
and with that of Girardi et al. (1998) (open squares)
and with a pure CDM model with $\Omega_0=1$, $h=1/2$ (dashed line).
Bahcall \& Cen (1992) estimated the cluster mass function using optical
data (richness, velocities, luminosity function of galaxies in clusters) 
as well as X-ray data (temperature distribution function of clusters).
Groups poorer than Abell clusters have also been included thus
extending the mass function to lower masses than the richer Abell clusters. 
Girardi et al. (1998) data are obtained from a sample of 152 nearby ($z \le 0.15$)
Abell-ACO clusters.
As shown, the CDM model that does not take account of the non-radial
motions over-produces the clusters abundance. The result is in agreement
with the study of the mass function in the SCDM model by Jing \& Fang (1994)
and by Bahcall \& Cen (1992a,b). Even with a lower normalization, CDM cannot
reproduce the cluster abundance as stressed by Bartlett \& Silk (1993), 
on the contrary a reduction of normalization produces a too steep
mass function (Bahcall \& Cen 1992b; Bartlett 1997). 
The introduction of non-radial motions (solid line) reduces
remarkably the abundance of clusters
with the result that
the model
predictions are in good agreement with the observational data.
This result confirms what found in Del Popolo \& Gambera (1999) showing
how a mass dependent threshold, $\delta_{\rm c}(M)$,
(dependence caused by the developing of non-radial motions) can solve several of
SCDM discrepancies with observations.

\section{Non-radial motions and the velocity dispersion function}

The VDF is defined in a similar way
to the mass function, namely it is the number of objects per unit volume 
with velocity dispersion larger than $\sigma_{\rm v}$. Since the velocity 
dispersion $\sigma_{\rm v}$ can be observed directly (on the contrary, 
mass measurement is usually model dependent), VDF provides a good 
test of theoretical models. Observed $\sigma_{\rm v}$ comes from the measurement of
galaxy redshift. The VDF can be calculated starting from the mass function:
\begin{equation}
n(\sigma_{\rm v}) =n(M) \frac{{\rm d} M}{{\rm d} \sigma_{\rm v}}
\label{eq:der}
\end{equation} 
The cumulative VDF can be obtained integrating Eq. (\ref{eq:der}):
\begin{equation}
n(>\sigma_{\rm v}) = \int_{\sigma_{\rm v}}^\infty n(\sigma_{\rm v'})
{\rm d} \sigma_{\rm v'}
\label{eq:derr}
\end{equation} 
In order to use Eq. (\ref{eq:der}) to calculate the VDF we need a 
relation between the velocity dispersion, $\sigma_{\rm v}$, and mass, $M$.
This can be obtained in several ways. 
The typical virial temperature may be written as:
\begin{equation}
k T_{\rm vir} =\frac{\alpha G \mu m_{\rm H}}{3} \frac{M_{\rm vir}}{R_{\rm vir}}
\label{eq:der1}
\end{equation}
where $\alpha$ is a factor of order unity. Evrard (1989) found in N-body 
simulations of a CDM model that typical clusters had $\alpha \simeq 0.8$. 
The molecular weight, $\mu$, corresponding to a fully 
ionized gas with primordial abundances is $\simeq 0.6$ and $m_{\rm H}$ is the
proton mass.
$M_{\rm vir}$ and $R_{\rm vir}$ are respectively the virial mass and virial 
radius, which are connected by:
\begin{equation}
M_{\rm vir}=178 \frac{4 \pi}{3} \rho_{\rm b} R_{\rm vir}^3
\end{equation}
Eq. (\ref{eq:der1}) has been tested in several N-body simulations. 
These numerical simulations give (Evrard 1996):
\begin{equation}
T_{\rm vir}=(6.8 h^{2/3} {\rm keV}) M^{2/3}
\label{eq:tt}
\end{equation}
The relation is so good that Evrard (1997) uses it as a primary 
mass indicator for observed clusters when considering the baryon 
fraction over an ensemble of clusters. 
Using:
\begin{equation}
\frac{3}{2} k T_{\rm vir} = \frac{1}{2} \mu m_{\rm H} \sigma_v^2
\end{equation}
(see Thomas \& Couchman 1992; Evrard 1990) 
togheter with Eq. (\ref{eq:tt}) we find the necessary relation 
between $\sigma_v$ and $M$:
\begin{equation}
\sigma_{\rm v}=824 {\rm km/s} \left(\frac{h M}{10^{15} M_{\odot}}\right)^{1/3}
\label{eq:ev}
\end{equation}
(Evrard 1989; Lilje 1990). 
\begin{figure}[ht]
\psfig{file=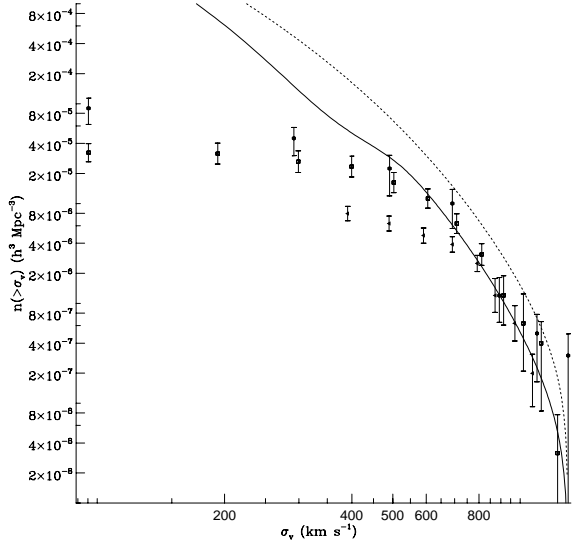,width=8.cm}
\caption{Cumulative VDF calculated using a CDM model without
taking into account non-radial motions (dashed line) and taking into account
non-radial motions (solid line) compared with Zabludoff et al. (1993) data (full dots)
and
with those by Mazure et al.(1996) (full triangles) for
$R \ge 1$ clusters and Fadda et al. (1996)
(open squares) for $R \ge -1$ clusters.
The theoretical curves are obtained using a $\sigma_{\rm v}$-$M$ relation with
zero scatter.}
\end{figure}
In Fig. 2 we compare the VDF obtained from a CDM model taking
account of non-radial motions (solid line) with the CfA 
data by Zabludoff et al. (1993) (full dots)
based on their survey of $R \geq 1$
Abell clusters within $z \leq 0.05$ and with the data by
Mazure et al. (1996) (full triangles) and Fadda et al. (1996) (open squares).
Mazure et al. (1996) data are obtained from a volume-limited
sample of 128 $R_{\rm ACO} \ge 1$ clusters
while that of Fadda et al. (1996)
are obtained from a sample of 172 nearby galaxy
clusters ($z \le 0.15$). 
We also plot the
VDF obtained from a CDM model without non-radial motions (dashed line). 
The SCDM model predicts more clusters than the CfA observation 
except at $\sigma_{\rm v} \simeq 1100  {\rm km/s}$. As reported by 
Jing \& Fang (1994) the SCDM model can be rejected at a very high 
confidence level ($>6 \sigma$). 
The reduction of
the normalization reduces the formation of clusters, thus resolving 
the problem of too many clusters, but leads to a deficit at 
$\sigma_v \simeq 1100  {\rm km/s}$.
When non-radial motions are taken into account (solid line) we obtain
a good agreement between theoretical predictions and observations. 
Both CDM and CDM with non-radial motions
predict more clusters of low velocity dispersion 
($\sigma_v \leq 300  {\rm km/s}$) than the observation. This discrepancy 
is not significant because the data at $\sigma_{\rm v}\leq 300 {\rm km/s}$ 
could be seriuosly underestimated because:
\begin{itemize}
\item as Zabludoff et al. (1993) stressed, their calculations of 
group velocity dispersions $\leq 300  {\rm km/s}$ are often underestimates; 
\item Zabludoff et al. (1993) measure $\sigma_{\rm v}$ only for CfA groups 
with $ \geq 5$ group members while for $\sigma_{\rm v} \leq 300  {\rm km/s}$ 
a fraction of groups could have less than 5 members. 
\end{itemize}
Also Fadda et al. (1996) 
cannot draw firm conclusions about their
incompleteness level, and hence about the behaviour of the $\sigma$
distribution for $\sigma \le 650  {\rm km/s}$, while Mazure et al. (1996)
has a supposed completeness limit of $\sigma \simeq 800  {\rm km/s}$.
A comparison between Eq. (\ref{eq:ev}) and N-body simulations of clusters
of galaxies in a CDM model shows that Eq. (\ref{eq:ev}) holds with a
rms scatter of $\simeq 10 \%$ (Evrard 1989; Lilje 1990).
To take into account the scatter in Eq. (\ref{eq:tt}) and Eq. (\ref{eq:ev})
it should be necessary to convolve $n(\sigma_{\rm v})$ with a Gaussian having
dispersion of $10\% \div 20\%$ (see Lilje 1990). The result of this operation
is that the abundance of clusters with high velocity dispersion
depends on the
assumed value of the rms scatter: a larger scatter implies a larger
$n(\sigma_{\rm v})$.
The result of this effect is the worsening of the problem of clusters
abundance over-production in the CDM model without non-radial motions.
 
\section{Non-radial motions and the shape of clusters}

Most clusters, like elliptical galaxies, are not spherical and their
shape is not due to rotation (Rood et al. 1972; Gregory \& Tifft 1976;
Dressler 1981). The perturbations that gave rise to the formation of
clusters of galaxies are alike to have been initially aspherical 
(Barrow \& Silk 1981; Peacock \& Heavens 1985; Bardeen et al. 1986) and
asphericities are then amplified during gravitational collapse (Lin
et al. 1965; Icke 1973; Barrow \& Silk 1981). The elongations are
probably due to a velocity anisotropy of the galaxies
(Aarseth \& Binney 1978) and according to Binney \& Silk (1979) and to 
Salvador-Sol\'e \& Solanes (1993) the elongation of clusters originates
in the tidal distortion by neighboring protoclusters. In particular
Salvador-Sol\'e \& Solanes (1993) found that the main distorsion on
a cluster is produced by the nearest neighboring cluster having more than 45
galaxies and the same model can explain the alignement between neighboring
clusters (Binggeli 1982; Oort 1983; Rhee \& Katgert 1987; Plionis 1993) and
that between clusters and their first ranked galaxy 
(Carter \& Metcalfe 1980; Dressler 1981; Binggeli 1982; Rhee \& Katgert 1987;
Tucker \& Peterson 1988; van Kampen \& Rhee 1990; Lambas et al. 1990;
West 1994). Clusters elongations and alignement
could be also explained by means of Zeldovich's (1978) "pancakes" theory
of cluster formation but this top-down formation model
is probably ruled out for several well known reasons (Peebles 1993). \\
The observational information on the
distribution of clusters shapes is sometimes conflicting.
Rhee et al. (1989) found that most clusters are
nearly spherical with ellipticities distribution having a peak
at $\epsilon \simeq 0.15$ while Carter \& Metcalfe (1980), Binggeli (1982),
Plionis et al. (1991) found that clusters are more elongated
with the peak of the ellipicities distribution at $\epsilon \simeq 0.5$.
More recently de Theije et al. (1995), de Theije et al. (1997) re-analyzed
the data studied by Rhee et al. (1989) and that by 
Plionis et al. (1991) concluding that:
\begin{description}
\item a) richer clusters are intrinsically more spherical than poorer ones; 
\item b) the projected elongations of clusters are consistent with a
prolate distribution
with clusters ellipticity distribution having a peak at $\epsilon
\simeq 0.4$
and extending to $\epsilon =0.8$;
\item c) in a $\Omega_{\rm 0} = 1$ CDM scenario clusters
tend to be less spherical than those
in a $\Omega_{\rm 0} = 0.2$ universe
and are too elongated with respect
to real observed clusters.
\end{description}
To study the effect of non-radial motions on the shape of clusters
we shall use a model introduced by Binney \& Silk (1979).
In that paper they showed that tidal interactions between protoclusters and the
neighbouring protostructures should yield prolate shapes (before virialization)
with an
axial ratio of protostructures of $\simeq 0.5 $, the typical value found in
clusters. After virialization the pre-existing elongation is damped
and the axial ratio leads to values of about
$0.7 \div 0.8$, that are higher with respect to observations. As observed by
Salvador-Sol\'e \& Solanes (1993) this last discrepancy  can be
removed taking into account that tidal interaction keeps going on after
virialization and that on average the damping of elongations
due to violent relaxation is eliminated by its growth after
virialization. Then this growth restores a value of $\epsilon$
near the one that clusters had before virialization.\\
\begin{figure}
\psfig{file=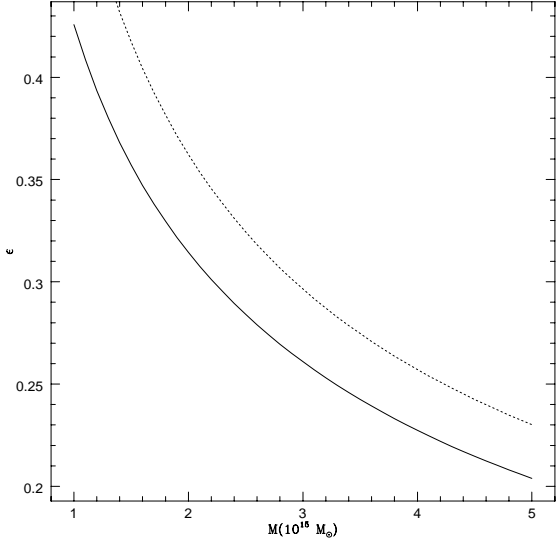,width=8.0cm}
\caption{Ellipticity, $\epsilon$, of clusters versus mass, $M$. The dashed
and solid lines
represent $\epsilon$ for a CDM without and with non-radial motions,
respectively.}
\end{figure}
According to the quoted Binney \& Silk (1979) model, an initially
spherical protostructure (e.g. a protocluster)
of mass $M$ having at distance $r(t)$ from its
centre a series of similar protostructure of mass $M'$ 
shall be distorted.
In order to calculate the distortion we must write the equation of
motion of a particle at position ${\bf {\it R}}$ relative to the centre of M
(assumed as origin of coordinates). Supposing that the effective perturbing
mass is less by a factor $\Delta=(\rho-\rho_{\rm b})/\rho_{\rm b}$ than its
true mass M',
in the limit
$|{\bf {\it r}}|>>|{\bf {\it R}}|$ we have:
\begin{equation}
{\bf \ddot {\it R}}=-\frac{G M}{R^3} {\bf {\it R}}+\frac{G \Delta M'}{r^3}
\left(
3 \frac{{\bf {\it R}} {\bf {\it r}}}{r^2} {\bf {\it r}} - {\bf {\it R}}
\right)
\end{equation}
If the tidal interaction is treated as first order,
writing ${\bf {\it R(t)}}={\bf {\it R_0}}+{\bf {\it R_1}}$
with ${\bf \ddot {\it R_0}}=-\frac{G M}{R_0^3} {\bf {\it R_0}}$ and
${\bf {\it R_0}}={\bf {\it R_m}} a(t)$,  
it is possible to show that the component of ${\bf {\it R_1}}$ parallel to
${\bf {\it R_m}}$ is:
\begin{equation}
R_1(x)=\frac{\mu}{2} \left[\frac{
3 ({\bf {\it r_m}} {\bf {\it R_m}})^2/r^2_{\rm m}-
R^2_{\rm m}}{R_{\rm m}}\right] G(x,x_1)
\end{equation}
where $x=2 a$ and $x_1$ is the value of $x$ at which the perturbation
was switched on.
The term in square parentheses reduces to $2 |R_{\rm m}|$, if
${\bf {\it R_m}}$ is parallel or antiparallel to ${\bf r_{\rm m}}$, and
to -$|R_{\rm m}|$ when they are orthogonal. Then the initially (at time
given by $x=x_1$)
spherical
density enhancement M becomes a prolate spheroid having ellipticity:
\begin{equation}
\epsilon=(1-b/a)=\frac{3}{2} \mu G(x,x_1)/(1+\mu G)
\simeq \frac{3}{2} \langle \mu^2 \rangle^{1/2} G(2,x_1)
\end{equation}
where $G(2,x_1)$ is defined in the quoted paper (see Eq. 13c) and
$\langle \mu^2 \rangle^{1/2}$ is given by (Binney \& Silk 1979): 
\begin{equation}
\langle \mu^2 \rangle^{1/2} = \left[\int \int \mu^2(M,M',r) N(M',r) {\rm d}r {\rm d} M'\right]^{1/2} 
\end{equation}
where
\[
\begin{array}{l}
\mu = \frac{\pi^2}{8} \left(\frac{R_{\rm m}}{r_{\rm m}}
\right)^3 \frac{M'}{M}\\
R_{\rm m} = \left(\frac{3 M}{4 \pi \rho_{\rm m}}\right)^{1/3}\\
r_{\rm m} = r(t_{\rm m})\\
\end{array}
\]
being $t_{\rm m}$ the time of maximum expansion, $r_{\rm m}=r(t_{\rm m})$,
and
$N(M',r) {\rm d}r$ is the number of condensations of mass $M'$ lying
between $r$ and $r+{\rm d}r$ from $M$. We calculated this quantity
using the
Press-Schechter's theory (see Sect. ~2):
\begin{equation}
N(M',r)=4 \pi r^2 n(M')
\end{equation}
To calculate $n(M')$ in the case of CDM without non-radial motions
we used Eq. (\ref{eq:press1}) while for a CDM with non-radial motions we used
Eq. (\ref{eq:press2}).
With the previous definitions $\langle \mu^2 \rangle^{1/2}$ is given by:
\begin{equation}
\langle \mu^2 \rangle^{1/2} = \frac{1}{9 \rho_{\rm b}^{1/2}}
\left[\int \frac{N(M',r) {\rm d}M'}{M+M'}\right]^{1/2}
\end{equation}
In Fig. 3 we show the shape of $\epsilon$ as a function of mass, $M$. \\
As shown $\epsilon$
declines with mass in agreement with the above quoted point a):
richer clusters are more spherical than poorer clusters. The physical
reason for this may be that regions of higher density turn
around earlier from Hubble flow than lower density regions (de Theije et al. 1995;
Ryden 1995).
The fundamental point in which we are interested is the effect of non-radial
motions on $\epsilon$. 
In a CDM model that takes
into account non-radial motions (solid line) $\epsilon$ is smaller than
in the simple CDM (dashed line). This is in agreement with the point c):
non-radial motions reduce the elongation of clusters. For a cluster of
$10^{15} M_{\odot}$ we get a value of $\epsilon \simeq 0.5$, if non-radial motions
are excluded, while $\epsilon \simeq 0.43$ when non-radial motions are taken
into account. Increasing the mass, as expected, clusters tend to become
more and more spherical. 
Finally Binney \& Silk (1979) model predicts that even a spherical density
enhancement $M$ soon becomes
a prolate spheroid in agreement with point b) and also with 
Salvador-Sol\'e \& Solanes (1993) result.

\section{Conclusions}

In this paper, using the model introduced by Del Popolo \& Gambera (1998a; 1999),
we have studied how non-radial motions change the mass function, the VDF and
the shape of clusters of galaxies.
We compared the theoretical mass function
obtained from the CDM model taking into account non-radial motions
with the experimental data by Bahcall \& Cen (1992a,b)
and Girardi et al. (1998).
The VDF, calculated
similarly to the mass function, was compared with the CfA data by 
Zabloudoff et al. (1993)
and those of Mazure et al. (1996) and
Fadda et al. (1996). 
Taking account of
non-radial motions we obtained a notheworthy reduction of
the discrepancies between the CDM predicted 
mass function, the VDF and the observations. Non-radial motions are also
able to change the shape of clusters of galaxies reducing their
elongations with respect to the prediction of the SCDM model.
This last result is in agreement with recent studies of the shapes of
clusters by de Theije et al. (1995, 1997).


\begin{flushleft}
{\it Acknowledgements}
We are grateful to an unknown referee for his helpful suggestions
and to Prof. E. Recami and Prof. E. Spedicato for stimulating
discussions while this work was in process.
\end{flushleft}


\begin{thebibliography}{}
\bibitem{} Aarseth S.J., Binney J., 1978, MNRAS 185, 227
\bibitem{} Bahcall N.A., Cen Y., 1992a, ApJ 398, L81
\bibitem{} Bahcall N.A., Cen Y., 1992b preprint
\bibitem{} Bahcall N.A., Soneira R.M., 1983, ApJ 270, 20
\bibitem{} Banday A.J., Gorski K.M., Bennet C.L., et al., 1996, ApJ COBE preprint 96-09, see also Sissa preprint astro-ph/9601065
\bibitem{} Bardeen J.M., Bond J.R., Kaiser N., Szalay A.S., 1986, ApJ 304, 15 \bibitem{} Barrow, J.D., Silk, J., 1981, ApJ 250, 432
\bibitem{} Bartlett J.G., 1997, Sissa preprint, astro-ph/9703090
\bibitem{} Bartlett J.G., Silk J., 1993, ApJ 407, L45
\bibitem{} Bennet C.L., Banday A., Gorski K.M., et al., 1996, ApJ 464, L1
\bibitem{} Binggeli B., 1982, A\&A 107, 338
\bibitem{} Binney J., Silk J., 1979, MNRAS 188, 273
\bibitem{} Blanchard A., Valls-Gabaud D., Mamon G., 1992, A\&A 264, 365
\bibitem{} Bond J.R., Szalay A.S., Silk J., 1988, ApJ 324, 627
\bibitem{} Bond J.R., Cole S., Efstathiou G., Kaiser N., 1991, ApJ 379, 440
\bibitem{} Bower R.G., Coles P., Frenk C.S., White S.D.M., 1993, ApJ 405, 403
\bibitem{} Brainerd T.G., Villumsen J.V., 1992, ApJ 394, 409
\bibitem{} Carter D., Metcalfe N., 1980, MNRAS 191, 325
\bibitem{} Cen R.Y., Gnedin N.Y., Kofman L.A., Ostriker J.P., 1992 preprint
\bibitem{} Colafrancesco S., Vittorio N., 1994, ApJ 422,443  
\bibitem{} Cole S., 1991, ApJ 367, 45
\bibitem{} Crone M.M., Evrard A.E., Richstone D.O., 1994, ApJ 434, 402
\bibitem{} Davis M., Efstathiou G., Frenk C.S., White S.D.M., 1985, ApJ 292, 371
\bibitem{} Dekel A., Bertshinger E., Yahil A. et al. 1992,  IRAS galaxies verses POTENT mass: density fields, biasing and $ \Omega$, Princeton preprint IASSNS-AST 92/55
\bibitem{} Del Popolo A., Gambera M., 1998a, A\&A 337, 96 
\bibitem{} Del Popolo A., Gambera M., 1998b, Proceedings of the VIII Conference on Theoretical Physics: General Relativity and Gravitation - Bistritza -  June 15-18, 1998 - Rumania
\bibitem{} Del Popolo A., Gambera M., 1999, A\&A 344, 17 (see also SISSA preprint astro-ph/9806044)
\bibitem{} de Theije P.A.M., Katgert P., van Kampen E., 1995, MNRAS 273, 30
\bibitem{} de Theije P.A.M., van Kampen E., Slijkhuis R.G., SISSA preprint 9705205 
\bibitem{} Dolgov A.D., 1997, SISSA preprint hep-ph/9707419
\bibitem{} Dressler A., 1978, ApJ 243, 26
\bibitem{} Edge A.C., Stewart G.C., Fabian A.C., Arnaud K.A., 1990, MNRAS 245, 559
\bibitem{} Efstathiou G., 1990, in ''The physics of the early Universe'', eds Heavens A., Peacock J., Davies A., (SUSSP)
\bibitem{} Efstathiou G., Sutherland W.J., Maddox S.J., 1990a, Nat., 348, 705
\bibitem{} Efstathiou G., Kaiser N., Saunders W. et al. 1990b, MNRAS 247, 10p
\bibitem{} Evrard A.E., 1989, ApJ 341, L71
\bibitem{} Evrard A.E., 1990, ApJ 363, 349
\bibitem{} Evrard A.E., 1997, Sissa Preprint astro-ph/9701148
\bibitem{} Evrard A.E., Metzler C.A., Navarro J.F., 1996, ApJ 469, 494
\bibitem{} Fadda D., Girardi M., Giuricin G., Mardirossian F., Mezzetti M., 1996, preprint SISSA astro-ph/9606098
\bibitem{} Mazure A., Katgert P., den Hartog R., et al. 1996, A\&A 311, 95 
\bibitem{} Frenk C.S., White S.D.M., Davis M., Efstathiou G., 1988 ApJ 327, 507
\bibitem{} Girardi M., Borgani S., Giuricin G., Mardirossian F., Mezzetti M., 1998, preprint SISSA astro-ph/9804188
\bibitem{} Gorsky K.M., Banday A., Bennet C.L., et al., 1996, ApJ COBE preprint 96-03, see also SISSA preprint astro-ph/9601063
\bibitem{} Gregory S.A., Tifft W.G., 1976, ApJ 205, 716 
\bibitem{} Henry J.P., Arnaud K.A., 1991, ApJ 372,410
\bibitem{} Hinshaw G., Banday A., Bennet C.L., et al., 1996, ApJ COBE preprint 96-04, see also SISSA preprint astro-ph/9601058
\bibitem{} Hoffman Y., Shaham J., 1985, ApJ 297, 16 
\bibitem{} Holtzman J., 1989, ApJS, 71, 1
\bibitem{} Holtzman J., Primack J., 1993, Phys. Rev. D43, 3155
\bibitem{} Icke V., 1973, A\&A 27, 1
\bibitem{} Jing Y.P., Fang L.Z., 1994, Phys. Rev. Lett. 73, 1882
\bibitem{} Jing Y.P., Mo H.J., Boerner G., Fang L.Z., 1994, A\&A 284, 703
\bibitem{} Kaiser N., Efstathiou G., Ellis R. et al. 1991, MNRAS 252, 1
\bibitem{} Kernan P.J., Sarkar S., 1996, Phys. Rev. D54, R3681 
\bibitem{} Klypin A., Holtzman J., Primack J., Reg\"os E., 1993, SISSA preprint astro-ph/9305011
\bibitem{} Lacey C., Cole S., 1993, MNRAS 262, 627
\bibitem{} Lacey C., Cole S., 1994, MNRAS 271, 676
\bibitem{} Lahav O., Edge A., Fabian A.C., Putney A., 1989, MNRAS 238, 881
\bibitem{} Lambas D.G., Nicotra M., Muriel H., Ruiz L., 1990, AJ 100, 1006
\bibitem{} Liddle A.R., Lyth D.H., 1993, Phys. Rev., 231, n 1, 2
\bibitem{} Lilje P.B., 1990, ApJ 351, 1
\bibitem{} Lilje P.B., 1992, ApJ 386, L33
\bibitem{} Lin C.C., Mestel L., Shu F.H., 1965, ApJ 142, 1431
\bibitem{} Maddox S.J., Efstathiou G., Sutherland W.J., Loveday J., 1990, MNRAS 242, 43p
\bibitem{} Monaco P., 1995, ApJ 447, 23
\bibitem{} Olive K.A., 1997, SISSA preprint astro-ph/9707212
\bibitem{} Oort J. H., 1983, ARA\&A 21, 373
\bibitem{} Ostriker J.P., 1993, ARA\&A 31, 689
\bibitem{} Peacock J.A., 1991, MNRAS 253, 1p
\bibitem{} Peacock J.A., Heavens A.F., 1985, MNRAS 217, 805
\bibitem{} Peacock J.A., Heavens A.F., 1990, MNRAS 243, 133
\bibitem{} Peacock J.A., Nicholson D., 1991, MNRAS 253, 307
\bibitem{} Peebles P.J.E., 1984, ApJ 284, 439
\bibitem{} Peebles P.J.E., 1993, Principles of Physical Cosmology, Princeton University Press
\bibitem{} Plionis M., 1993, SISSA preprint astro-ph/9312013
\bibitem{} Plionis M., Barrow J.D., Frenk C.S., 1991, MNRAS 249, 662
\bibitem{} Press W.H., Schechter P., 1974, ApJ 187, 425 
\bibitem{} Rhee G.F.R.N., Katgert P., 1987, A\&A 183, 217
\bibitem{} Rhee G.F.R.N., van Haarlem M.P., Katgert P., 1989, A\&AS 91, 513 
\bibitem{} Rood H.H., Page T.L., Kintner E.C., King I.R., 1972, ApJ 175, 627
\bibitem{} Ryden B.S., SISSA preprint astro-ph/9510105
\bibitem{} Salvador-Sol\'e E., Solanes J.M., 1993, ApJ 417, 427
\bibitem{} Saunders W., Frenk C., Rowan-Robinson M. et al., 1991, Nat 349, 32 
\bibitem{} Schaefer R.K., 1991, Int. J. Mod. Phys. A6, 2075
\bibitem{} Schaefer R.K., Shafi Q., 1993, Phys. Rev., D47, 1333
\bibitem{} Schaefer R.K., Shafi Q., Stecker F., 1989, ApJ 347, 575
\bibitem{} Shafi Q., Stecker F.W., 1984, Phys. Rev. D29, 187
\bibitem{} Smoot G.F., Bennett C.L., Kogut A., et al., 1992, ApJ 396, L1
\bibitem{} Steigman G., 1996, SISSA preprint astro-ph/9608084
\bibitem{} Thomas P.S., Couchman H.M.P., 1992, MNRAS 257, 11
\bibitem{} Tucker G.S., Peterson J.B., 1988, AJ 95, 298
\bibitem{} Turner M.S., 1991, Phys. Scr. 36, 167
\bibitem{} Valdarnini R., Bonometto S.A., 1985, A\&A 146, 235
\bibitem{} van Kampen E., Rhee G.F.R.N., 1990, A\&A 237, 283 
\bibitem{} West M.J., 1994, MNRAS 268, 79
\bibitem{} White S.D.M., Efstathiou G., Frenk C.S., 1993a, The amplitude of the mass fluctuations in the Universe, Durham preprint
\bibitem{} White S.D.M.., Efstathiou G., Frenk C.S., 1993b, MNRAS 262, 1023
\bibitem{} White S.D.M.., Frenk C.S., Davis M., Efstathiou G., 1987, ApJ 313, 505
\bibitem{} Zabludoff A.I., Geller M.J., Huchra J.P, Ramella M., 1993, AJ 106, 1301
\bibitem{} Zeldovich Ya.B., 1978, Longair M.S., Einasto J., Eds., Proc. IAU Symp. No. 79, The Large Scale Structure of the Universe, Dordrecht, p. 409
\end{thebibliography}
\end{document}